\documentclass[nohyper,12pt,letterpaper]{JHEP}
\usepackage{epsfig}
\newfont{\frak}{eufm10 scaled 1200}

\newfont{\Bbb}{msbm10 scaled 1200}
\newcommand{\mathbb}[1]{\mbox{\Bbb #1}}
\DeclareSymbolFont{AMSa}{U}{msa}{m}{n}
\DeclareSymbolFont{AMSb}{U}{msb}{m}{n}
\let\Box\relax
\DeclareMathSymbol{\Box}{\mathord}{AMSa}{"03}

\title{Running of the Spectral Index in Noncommutative Inflation}

\author{S. A. Alavi\\
Department of Physics, Tarbiat Moallem University of Sabzevar,\\
P.O.Box 397, Sabzevar, Iran\\
E-mail: \email{alavi@sttu.ac.ir}}

\author{Forough Nasseri\\
Department of Physics, Tarbiat Moallem University of Sabzevar,\\
P.O.Box 397, Sabzevar, Iran\\
and\\
Khayyam Planetarium, P.O.Box 844, Neishabour, Iran\\
E-mail: \email{nasseri@fastmail.fm}}

\abstract{We study the cosmological implications of the space-space
noncommutative inflation and present formulae for the
spectral index and its running.
Our results show that deviations from the spectral index and its
running depend on the space-space
noncommutativity length scale. We conclude that in the slow-roll regime
of a typical inflationary scenario, space-space noncommutativity has
negligible effects on both the spectral index and its running.
Two classes of examples have been studied and comparisons
made with the standard slow-roll formulae. The results show that the
correction terms to the commutative case are positive for the spectral
index and negative for the running of the spectral index.
Nevertheless the spectral index in noncommutative spaces
is less than one owing to the very small values of the correction terms.}

\keywords{Noncommutative Geometry;
Cosmology of Theories beyond the SM.\\
Report Number: STTU-Phys-01-2004}

\date{today}
\preprint{STTU-Phys-01-2004}
\begin{document}

\section{Introduction}
It is generally believed that the picture of space-time as a manifold
should break down at very short distances of the order of the Planck
length. Field theories on noncommutative spaces may play an important
role in unraveling the properties of nature at the Planck scale.
It has been shown that the noncommutative geometry naturally appears in
string theory with a non zero antisymmetric B-field [1].

Besides the string theory arguments the noncommutative field theories
by themselves are very interesting. In a noncommutative space-time
the coordinates operators satisfy the commutative relation
\begin{equation}
[{\hat x}^{\mu},{\hat x}^{\nu}]=i\theta^{\mu\nu},
\end{equation}
where $\hat x$ are the coordinate operators and
$\theta^{\mu\nu}$ is an antisymmetric tensor of dimension of 
(length)$^2$.
Generally noncommutative version of a field theory is obtained by replacing
the product of the fields appearing in the action by the star products
\begin{equation}
(f \star g)(x)=\exp \left( \frac{i}{2} \theta^{\mu\nu}
\frac{\partial}{\partial x^\mu}\frac{\partial}{\partial y^\nu}\right)
f(x)g(y) \mid_{x=y}
\end{equation}
where $f$ and $g$ are two arbitrary functions which we assume to be
infinitely differentiable.

In recent years there have been a lot of work devoted to the study
of noncommutative field theory or noncommutative quantum mechanics
and possible experimental consequences of extensions of the
standard formalism [2-13]. In the last few
years there has been also a growing interest in probing the space-space
noncommutativity effects on cosmological observations [14-26].

Apart from the field theory or quantum mechanics,
we are more interested in some possible cosmological consequences
of noncommutativity in space. In so doing,
we study the effects of space-space noncommutativity on
the spectral index and its running. This issue has been studied
in the last few years by several authors [21-26].

The theory of inflation [27] has faired well in this latest round
of cosmological discoveries. Generically, slow-roll inflation predicts
the Universe is flat, and that the primordial perturbations are
Gaussian, adiabatic, and have a nearly scale invariant spectrum.
The degree to which slow-roll inflation predicts a scale invariant
spectrum depends on the dynamics of the scalar field(s) controlling
inflation. The simplest possibility is a single `inflaton', slowing
rolling down its potential with its kinetic energy strongly damped
by the Hubble expansion. In the limit in which the rolling is infinitely
slow and the damping infinitely strong, the primordial spectrum is a
power law, with index $n$ exactly equal to one. Deviations from $n=1$ are
measures of how slowly the field rolled and how strongly its motion
was damped during inflation. Equivalently, different inflationary
models predict different values of $n$ or more generally of the shape
of the spectrum, measurements of this primordial spectrum enable one to
discriminate among different inflationary models.

There is another reason why precise measurements of the primordial
spectrum are important to proponents of inflation. Even before inflation
was proposed, Harrison and Zel'dovich introduced the notion that scale
free ($n=1$) adiabatic perturbations represent natural initial conditions.
A spectrum with $n$ not exactly equal to one, or even more telling, one
with deviations from a pure power law form, is perfectly compatible
with inflation. While not a {\it proof} of inflation, such deviations
would surely be a {\it disproof} of the Harrison-Zel'dovich speculations.

The spectral index and its running in commutative geometry have been
studied by several authors, for example [28-37].
In the slow-roll approximation, $|n-1|$ is small. It is often assumed
[30] that deviations from a pure power law are of order of $(n-1)^2$. If
true, this would mean that the recent measurements indicating $|n-1|$
is smaller than about $0.1$ imply that deviations from a power law would
only show up at the percent level at best.

We will use a natural unit system that sets $k_B$, $c$ and $\hbar$
all equal to $1$, so that $\ell_P=M_P^{-1}=\sqrt{G}$.

The plan of this article is as follows. In section 2, we give a
brief review of the spectral index and its running in the standard
and commutative inflation. In section 3, we first present the explicit
and general formulae of the spectral index and its running in
noncommutative spaces and then apply them to two classes of examples
of the inflaton potential. Finally, we discuss our results and
conclude in section 4.

\section{Running of the spectral index in commutative inflation}

In general, the power spectrum of the scalar perturbations is closely
related to the functional form of the inflaton potential, $V(\phi)$ [37]
\begin{equation}
\label{3}
\delta_H^2=\frac{128 \pi}{3 M_P^6}\frac{V^3}{V'^2},
\end{equation}
where $M_P$ is the Planck mass and a prime denotes
$d/d\phi$\footnote{We employ the normalization conventions of Ref.[31].}.
The relationship between the inflaton field and comoving wave-number
follows from the scalar field equations of motion and is given by
\begin{equation}
\label{4}
\frac{d}{d\ln k}=-\frac{M_P^2}{8\pi}\frac{V'}{V}\frac{d}{d\phi}
\end{equation}
in the slow-roll limit. By defining the slow-roll parameters as [29]
\begin{eqnarray}
\label{5}
\epsilon &\equiv& \frac{M_P^2}{16 \pi}\left( \frac{V'}{V} \right)^2,\\
\label{6}
\eta &\equiv& \frac{M_P^2}{8 \pi} \frac{V''}{V},\\
\label{7}
\xi &\equiv& \frac{M_P^4}{64 \pi^2} \frac{V' V'''}{V^2},
\end{eqnarray}
the spectral index and its running may be expressed directly in terms of
the potential and its derivatives [28,30]
\begin{eqnarray}
\label{8}
n_{\rm S}-1 &=& \frac{d \ln \delta_H^2}{d \ln k}=-6\epsilon+2 \eta,\\
\label{9}
\frac{d n_{\rm S}}{d \ln k} &=& 16 \epsilon \eta -24 \epsilon^2 -2\xi.
\end{eqnarray}
Then the running of the spectral index depends on the third derivative
of the potential. Eq. (\ref{9}) is truncated at order $\xi$, such that
quadratic corrections in $\epsilon$ and $\eta$ are assumed to be
negligible. This requires that $|\xi| \ll {\rm {max}} (\epsilon,
|\eta|)$ and is equivalent to assuming that $|n_S-1| \ll 1$ and
$|dn_S/d\ln k| \approx (n_S-1)^2$ or less. As emphasized in Refs.
[34,35], slow-roll predicts the former condition but not necessarily
the latter.

\section{Running and space-space noncommutativity}

For a polynomial potential the {\it effective} action will be of
the form [22]
\begin{eqnarray}
\label{10}
S&=&\int d^4 x \sqrt{-g} \bigg[ \frac{1}{2} (\partial_{\mu}\phi)
\star (\partial^{\mu}\phi)-\frac{\lambda}{n!}\phi\star...\star\phi
\bigg]\nonumber\\
&\equiv& \int d^4x \sqrt{-g}\bigg[\frac{1}{2}(\partial_{\mu}\phi)
\star(\partial^{\mu}\phi)-\frac{\lambda}{n!}\phi^{\star n}\bigg].
\end{eqnarray}

With no loss of generality, the authors of Ref. [22] choose a frame in
which the only nonvanishing space-space component of $\theta^{\mu\nu}$ is
\begin{equation}
\label{11}
\theta^{12}(t)=\frac{1}{\Lambda^2 a^2},
\end{equation}
where $a$ is the scale factor of the Universe and $\Lambda^{-1}$ the
noncommutativity length scale. Using $\vartheta$ as the angle
between the comoving wave-number $k$ and the third axis, the scalar
fluctuations have the late time amplitude [22]
\begin{equation}
\label{12}
|\phi_{\rm k}|=\frac{H}{\sqrt{2} k^{3/2}}
\left( 1-\frac{3}{32} \frac{H^4}{\Lambda^4}\sin^2{\vartheta} \right).
\end{equation}
The first term in (\ref{12}) is the standard result, valid for
fluctuations of a massless field in a de Sitter Universe, see Ref. [38].
The second is instead a new effect induced by $\theta^{\mu\nu}$
depending terms in the action (\ref{10}), and explicitly shows the
presence of a preferred direction associated with the nonvanishing
component $\theta^{12}$. It is worth noticing that this correction does
not decrease at later times, so that the imprint of noncommutativity
is preserved even after the physical scales of the fluctuations have
grown to much larger sizes than $\Lambda^{-1}$. Our interested purpose
here is actually determination of both the spectral index and its running in
space-space noncommutativity, but this was not done in [22].

In the commutative or usual inflation, the density perturbations can be
expressed by [38]
\begin{equation}
\label{13}
\delta_H^2=\frac{4\pi k^3}{(2\pi)^3}|\phi_k|^2\left(
\frac{H}{\dot \phi} \right)^2.
\end{equation}
Using (\ref{12}) and (\ref{13}), we have
\begin{equation}
\label{14}
\delta_H^2=\left(\frac{H}{2\pi}\right)^2 \left( 1-\frac{3}{32} \frac{H^4}
{\Lambda^4} \sin^2\vartheta \right)^2 \left( \frac{H}{\dot{\phi}} \right)^2.
\end{equation}
Based on the slow-roll approximation, the equations of motion take
the forms [38]
\begin{eqnarray}
\label{15}
H^2 &=& \frac{8 \pi}{3 M_P^2} V,\\
\label{16}
3 H \dot{\phi} &=& - V',
\end{eqnarray}
where we assume the Universe to be spatially flat and the inflaton field
$\phi$ to be spatially homogeneous.
From (\ref{15}) and (\ref{16}), for space-space noncommutative geometry
the density perturbation is
\begin{equation}
\label{17}
\delta_H^2=\frac{128 \pi}{3 M_P^6}
\frac{V^3}{V'^2} \left( 1-\frac{3}{32}\frac{H^4}{\Lambda^4} \sin^2
\vartheta\right)^2.
\end{equation}
When $\Lambda \to +\infty$ or $\Lambda^{-1}=0$, 
Eq.(\ref{17}) reproduces the standard and commutative density
perturbation.
Substituting (\ref{17}) in (\ref{8}) and (\ref{9}), one can
easily obtain the formulae for the spectral index and its running
\begin{eqnarray}
\label{18}
n_{\rm S}-1&=&-6\epsilon+2\eta+\frac{\pi \sin^2 \vartheta}{3 M_P^2 \Lambda^4}
V'^2\left( 1-\frac{3}{32}\frac{H^4}{\Lambda^4} \sin^2\vartheta
\right)^{-1},\\
\label{19}
\frac{dn_{\rm S}}{d \ln k}&=& 16 \epsilon \eta -24 \epsilon^2-2 \xi
-\frac{\sin^2\vartheta}{3 \Lambda^4} \Bigg( 1 - \frac{3}{32}
\frac{H^4}{\Lambda^4} {\sin^2\vartheta} \Bigg)^{-2}\nonumber\\
&\times&\Bigg[ \frac{V'^2 V''}{4V}\Bigg( 1-\frac{3}{32}
\frac{H^4}{\Lambda^4} \sin^2\vartheta \Bigg)
+\frac{\pi^2 V'^4 \sin^2\vartheta}{6 M_P^4 \Lambda^4}\Bigg].
\end{eqnarray}
The point is that the correction term to the spectral index which is
the third term of the RHS of (\ref{18}) and to the running of the
spectral index which is the fourth term of the RHS of (\ref{19})
have positive and negative values, respectively.
The sum of the first and second term in the RHS of (\ref{18}) is
negative\footnote{See the first term of the RHS of (\ref{26})
and (\ref{36}) which have negative values.}.
This makes $n_S-1$ to have negative value and so
$n_S$ to be a few smaller than one in commutative inflation.
Due to very small value of $\Lambda^{-1}$, being of the
order of the Planck length, the correction terms to the commutative
spectral index is much smaller than $-6\epsilon+2\eta$, see Eq.(\ref{18}).
Consequently, the value of $n_S-1$ is negative and so $n_S$ is a few
smaller than one in noncommutative inflation.
Let us now use (\ref{18}) and (\ref{19}) for two 
potentials of the inflation. We first study $V(\phi)=m^2\phi^2/2$
and then $V(\phi)=\lambda \phi^4$.
In the limit of $\Lambda \to +\infty$, (\ref{18}) and (\ref{19})
approach to (\ref{8}) and (\ref{9}), respectively.

\subsection{The first example}
For the purpose of illustration, we now consider the potential 
$V(\phi)=m^2\phi^2/2$

\begin{eqnarray}
\label{20}
\epsilon&=&\eta=\frac{M_P^2}{4\pi \phi^2},\\
\label{21}
\xi&=&0.
\end{eqnarray}
Using the definition of e-folding in inflation [38]
\begin{equation}
\label{22}
N=-\frac{8\pi}{M_P^2}\int^{\phi_{\rm f}}_{\phi}\frac{V}{V'}d\phi,
\end{equation}
we have
\begin{equation}
\label{23}
N=\frac{2\pi}{M_P^2}(\phi^2-\phi^2_{\rm f}),
\end{equation}
where $\phi_{\rm f}$ is the value of the inflaton field at the
end of inflation.
To obtain e-folding as a function of the inflaton field,
we must obtain $\phi_{\rm f}$. From $\epsilon=\eta=1$
we get
\begin{equation}
\label{24}
\phi_{\rm f}=\frac{M_P}{\sqrt{4\pi}}.
\end{equation}
From (\ref{23}) and (\ref{24}), we have
\begin{equation}
\label{25}
\phi^2=\frac{M_P^2}{4\pi}(2N+1).
\end{equation}
With no loss of generality, we take 
$\vartheta=\frac{\pi}{2}$. So substituting $\sin\vartheta=1$
in the above equations leads us to
\begin{eqnarray}
\label{26}
n_{\rm S}-1&=&-\frac{4}{(2N+1)}+\frac{(2N+1)m^4}{12\Lambda^4}
\bigg[ 1- \frac{(2N+1)^2m^4}{96\Lambda^4} \bigg]^{-1},\\
\label{27}
\frac{dn_{\rm S}}{d \ln k}&=&-\frac{8}{(2N+1)^2}-
\frac{m^4}{6 \Lambda^4}\bigg( 1-\frac{(2N+1)^2m^4}{96\Lambda^4}
\bigg)^{-2}\bigg[ 1+ \frac{(2N+1)^2m^4}{96 \Lambda^4} \bigg].
\end{eqnarray}
The point is that the general formulae of (\ref{26}) and
(\ref{27}) in terms of unknown value of $\vartheta$ could be obtained
by substituting $\Lambda^2$ with $\Lambda^2/\sin \vartheta$,
see (\ref{12}).

We know the mass of the inflaton field
to be $m\simeq 1.21 \times 10^{-6} M_P$,
so the term of $\frac{m^4 (2N+1)^2}{96 \Lambda^4}$ is much less
than one and we can rewrite (\ref{26}) and (\ref{27}) up to the
second order of $m^4/\Lambda^4$ for the spectral index

\begin{eqnarray}
\label{28}
n_{\rm S}-1&\simeq&-\frac{4}{(2N+1)}+\frac{(2N+1)m^4}{12\Lambda^4}
\bigg[ 1+ \frac{(2N+1)^2m^4}{96\Lambda^4} \bigg]+...\nonumber\\
&\simeq&-\frac{4}{(2N+1)}+\frac{(2N+1)m^4}{12\Lambda^4}
+\frac{(2N+1)^3m^8}{1152\Lambda^8}+...,
\end{eqnarray}
and up to the third order of $m^4/\Lambda^4$ for the running of the
spectral index
\begin{eqnarray}
\label{29}
\frac{dn_{\rm S}}{d \ln k}&\simeq&-\frac{8}{(2N+1)^2}-
\frac{m^4}{6 \Lambda^4}\bigg( 1+\frac{(2N+1)^2m^4}{48\Lambda^4}\bigg)
\bigg[ 1+ \frac{(2N+1)^2m^4}{96 \Lambda^4} \bigg]+...\nonumber\\
&\simeq&-\frac{8}{(2N+1)^2}-\frac{m^4}{6 \Lambda^4}-
\frac{(2N+1)^2m^8}{192\Lambda^8}
-\frac{(2N+1)^4m^{12}}{27648\Lambda^{12}}+....
\end{eqnarray}
These equations tell us that the effects of the space-space
noncommutativity on the spectral index and
its running are of the order of $m^4/\Lambda^4$,
$m^8/\Lambda^8$ and $m^{12}/\Lambda^{12}$ or so.
According to the value of e-folding to be approximately
$60$ for solving the problems of the standard cosmology,
and taking $\Lambda^{-1}$ to be of the order of the Planck length,
we conclude that the second, third and fourth terms in the 
RHS of (\ref{28}) and (\ref{29}) are much smaller than
the first term because $m/M_P \sim 10^{-6}$.
Therefore $m^4/\Lambda^4 \sim 10^{-24}$,
$m^8/\Lambda^8 \sim 10^{-48}$ and $m^{12}/\Lambda^{12} \sim 10^{-72}$.
Since the accuracy of the Wilkinson Microwave Anisotropy Probe
(WMAP) data of the spectral index
and its running is up to only two or three decimal integers,
for example $n_S= 0.99\pm 0.04$ [39] or $dn_S/d\ln k=-0.055^{+0.028}
_{-0.029}$ [40], we conclude that the effects of the noncommutativity
of spaces which are of the order of $m^4/\Lambda^4$ and its powers of
$2$ and $3$ or so cannot be currently detected by the present experimental
celestial or ground instruments. If true, the imprints of the
space-space noncommutative geometry on the spectral index and its
running are undetectable and negligible.

\subsection{The second example}
For the second example, we study $V(\phi)=\lambda \phi^4$.
The slow-roll parameters are
\begin{eqnarray}
\label{30}
\epsilon&=&\frac{M_P^2}{\pi \phi^2},\\
\label{31}
\eta&=&\frac{3 M_P^2}{2\pi \phi^2},\\
\label{32}
\xi&=&\frac{3 M_P^4}{2 \pi^2 \phi^4}.
\end{eqnarray}
Using the definition of e-folding in inflation, we have
\begin{equation}
\label{33}
N=\frac{\pi}{M_P^2}\bigg(\phi^2-\phi^2_{\rm f}\bigg).
\end{equation}
To obtain the e-folding number as a function of the inflaton field,
we must obtain the value of the inflaton field at the end of inflation,
$\phi_{\rm f}$. From $\epsilon=1$
we get
\begin{equation}
\label{34}
\phi_{\rm f}=\frac{M_P}{\sqrt{\pi}}.
\end{equation}
Assuming $\eta=1$ and $\xi=1$, we obtain
$\phi_{\rm f}=\sqrt{\frac{3}{2\pi}}M_P$ and
$\phi_{\rm f}=(\frac{3}{2\pi^2})^{1/4}M_P$,
respectively. These values of $\phi_{\rm f}$ based on the conditions
$\eta=1$ and $\xi=1$ are larger than $\phi_{\rm f}$ arisen from
$\epsilon=1$, see (\ref{34}). We here take the condition $\epsilon=1$
by itself is a true condition to obtain $\phi_{\rm f}$.
From (\ref{33}) and (\ref{34}) we
have
\begin{equation}
\label{35}
\phi^2=\frac{M_P^2(N+1)}{\pi}.
\end{equation}
Again we take $\vartheta=\frac{\pi}{2}$. Substituting 
$\sin\vartheta=1$ in (\ref{18}) and (\ref{19}) and using
(\ref{30}), (\ref{31}) and (\ref{32}) we obtain the main formulae
for the spectral index and its running 
\begin{eqnarray}
\label{36}
n_{\rm S}-1&=&-\frac{3}{(N+1)}+\frac{16\lambda^2M_P^4}{3\pi^2\Lambda^4}
(N+1)^3\bigg[ 1- \frac{2\lambda^2M_P^4(N+1)^4}{3\pi^2\Lambda^4}
\bigg]^{-1},\\
\label{37}
\frac{dn_{\rm S}}{d \ln k}&=&-\frac{3}{(N+1)^2}-
\frac{16 \lambda^2 M_P^4 (N+1)^2}{3 \pi^2 \Lambda^4}
\bigg(1-\frac{2\lambda^2M_P^4(N+1)^4}{3\pi^2\Lambda^4}\bigg)^{-2}\nonumber\\
&\times&\bigg[3+\frac{2\lambda^2M_P^4(N+1)^4}{3\pi^2\Lambda^4}\bigg].
\end{eqnarray}
Considering the noncommutativity length
scale $\Lambda^{-1}$ to be of the order of the Planck scale, so we obtain
$M_P^4/\Lambda^{4}$ of the order of unity, ${\cal{O}}(1)$.
Potentials in chaotic inflation are characterized by a small overall
coupling constant, $\lambda \simeq 10^{-15}$, to ensure consistency
with the amplitude of energy density perturbations around $10^{-5}$,
for $\phi$ field values of a few Planck units.
So we can expand (\ref{36}) and
(\ref{37}) in terms
of $\frac{2 \lambda^2 M_P^4 (N+1)^4}{3 \pi^2 \Lambda^4}$.
After a lengthy but straightforward calculation, we find
\begin{eqnarray}
\label{38}
n_{\rm S}-1&\simeq&-\frac{3}{(N+1)}+\frac{16\lambda^2M_P^4}{3\pi^2\Lambda^4}
(N+1)^3\bigg[ 1+ \frac{2\lambda^2M_P^4(N+1)^4}{3\pi^2\Lambda^4}
\bigg]+...\nonumber\\
&\simeq&-\frac{3}{(N+1)}+\frac{16\lambda^2M_P^4(N+1)^3}{3\pi^2\Lambda^4}
+\frac{32\lambda^4M_P^8(N+1)^7}{9\pi^4\Lambda^8}+...,\\
\label{39}
\frac{dn_{\rm S}}{d \ln k}&=&-\frac{3}{(N+1)^2}-
\frac{16 \lambda^2 M_P^4 (N+1)^2}{3 \pi^2 \Lambda^4}
\bigg(1+\frac{4\lambda^2M_P^4(N+1)^4}{3\pi^2\Lambda^4}\bigg)\nonumber\\
&\times&
\bigg[3+\frac{2\lambda^2M_P^4(N+1)^4}{3\pi^2\Lambda^4}\bigg]
+...\nonumber\\
&\simeq&-\frac{3}{(N+1)^2}-\frac{48\lambda^2M_P^4(N+1)^2}{3\pi^2
\Lambda^4}-\frac{224\lambda^4M_P^8(N+1)^6}{9\pi^4\Lambda^8}\nonumber\\
&-&\frac{128\lambda^6M_P^{12}(N+1)^{10}}{27\pi^6\Lambda^{12}}+...,
\end{eqnarray}
where we have truncated at the second order
of $\lambda^2 M_P^4/\Lambda^4$ for the
spectral index, and also at the third order
of $\lambda^2 M_P^4/\Lambda^4$ for the
running of the spectral index.
Again similar to the case of $m^2 \phi^2/2$
we conclude that in comparison with the first terms in the RHS
of (\ref{38}) and (\ref{39}), the second, third and fourth terms
in RHS of mentioned equations can be ignored because 
$\lambda \sim 10^{-15}$ and the e-folding number to be approximately
close to 60. So, the imprints of the space-space noncommutative
geometry are negligible by considering the accuracy of the WMAP
data and of the present experimental celectial or ground instruments.
\section{Conclusions}

We have studied the effects of the space-space noncommutative
geometry on the spectral index and its running, and obtained
the corrections arisen from the space-space noncommutativity.
If there exists any noncommutativity of space
in nature, as it seems to emerge from different theories and
arguments, its implications should appear in the spectral index and
its running. We presented the general features of our formalism and
applied it to two specific potentials of chaotic inflation.
The results showed that the terms arising from the noncommutativity
of space depend on the space-space noncommutativity length scale.

For the purpose of illustration, we present the formulae
for the spectral index and its running in noncommutative
inflation for two classes of examples of the inflationary potentials.
The first one is $m^2\phi^2/2$ and the second $\lambda \phi^4$.
For the former we conclude that the space-space
noncommutativity effects on both the spectral index
and its running to be of the order of $m^4/\Lambda^4$ and its powers
of $2$, $3$ or so, and for the latter to be of the order of 
$\lambda^2M_P^4/\Lambda^4$ and its power of $2$, $3$ or so.
Our results are in terms
of the noncommutativity length scale $\Lambda^{-1}$, to be approximately
close to the Planck length. In the limit of $\Lambda \to +\infty$
or $\Lambda^{-1}=0$, our formalism approach to those of the usual or
commutative inflation [38].

Since the values of $m^4/\Lambda^4$ and $\lambda^2M_P^4/\Lambda^4$
are very small, the effects of space-space noncommutativity on
the spectral index and its running are very small.
Considering the accuracy of the WMAP data which is up to two or three
decimal integers as given in [39] and [40], we conclude that the effects
of the space-space noncommutative geometry on the spectral index
and its running cannot be currently detected.
This means that the noncommutativity
corrections, being of the order of $10^{-24}$ or so
for $V(\phi)=m^2\phi^2/2$ and $10^{-30}$ or so for
$V(\phi)=\lambda \phi^4$, are both negligible and undetectable
by the present experimental celectial or ground instruments. 
Similar to our results here, recent paper [19] shows that in the
slow-roll regime of a typical chaotic inflationary scenario,
noncommutativity of space has negligible impact.

In general, commutative inflationary cosmology satisfying the slow-roll
conditions naturally predicts positive spectral index which is a few
smaller than one
and also predicts negative and very small value for the running of the
spectral index. Our results
as given in (\ref{18}) and (\ref{19}) show that the correction
terms to the commutative case are positive for $n_S-1$ and
negative for the running of the spectral index. This means that the
space-space noncommutativity causes a negligible increasing in the value
of $n_S-1$ and  a negligible decreasing in the value of
$\frac{dn_S}{d \ln k}$. Nevertheless the spectral index in
the space-space noncommutative inflation is less than one owing to the
very small values of the correction terms. This result for the value of
the spectral index in noncommutative spaces is in agreement with the
WMAP data, for example $n_S=0.99\pm 0.04$ [39].

\section*{Acknowledgments}
It is a pleasure to thank Fedele Lizzi, Robert Brandenberger
and Richard Easther for helpful comments.
F.N. thanks Amir and Shahrokh for useful helps.

%\subsection*{ Appendix A}

\end{document}